# Future Mobile Network Architecture: Challenges and Issues

**Muhammad Bilal, Moonsoo Kang**
Dept. of Computer Eng., Chosun University
Gwangju, 501-759 South Korea
[e-mail: engr.mbilal@yahoo.com, mskang@chosun.ac.kr]
*Corresponding author: Moonsoo Kang*

*Abstract*—The future mobile networks facing many challenges and to cope these challenges, different standards and project has been proposed so far. Most recently Cognitive Networks has opened a new ground to present suitable architecture and mechanism for these challenges. The objective of this paper is to identify and discuss the challenges to the future mobile networks and to discuss some workable solutions to these challenges. Finally, on the basis of discussion a simple flexible network architecture is proposed.

## I. INTRODUCTION

The mobile communication started in early 1980, and from that time the mobile technology has been developed to fulfill the voice and data requirements of the subscriber. But the data requirements are increasing day to day because subscriber demands, fast mobile internet for online multimedia applications, video calls and video conferencing and also need fast data rate for business applications etc. Therefore, different wireless stander and technologies came in the market to enhance the data capacity of the networks and to provide new services and applications. All these systems are designed independently; targeting different service types and data rates. None of these standers meeting the public demands and hence incapable to provide new services and applications that are ubiquitous and customized to individual needs. Similarly, because of independent Radio Access Technologies (RATs) the radio spectrum is underutilized. Hence the operator is not getting the optimal net profit. In order to meet the rising expectations, all wired and wireless networks need to be integrated to share the resources and services. This convergence introduces the concept of Converged Heterogeneous Networks (CHN). Therefore, we need:

- Context aware and active mobile networks (MN).
- Intelligent, reconfigurable, cognitive and cooperative mobile terminal (MT).

By using context aware, active and cognitive MN and MT a very flexible networking environment can be established which can address the CHN issues. There is a lot of research has been carried out to deal with the future network challenges. Most of the research is more specific and targeting specific issues. There is a need to combine the ideas and give a complete network architectural solution for future networks. In the subsequent sections of this paper will discuss the important challenges to future networks and gives an architectural solution to those challenges.

## II. CHALLENGES TO FUTURE MOBILE NETWORKS

The main objective of future mobile networks is to integrate all mobile networks by providing the roaming and seamless handover facility between different cellular networks and public private unlicensed networks, hence raising different challenges.

We can divide the challenges of future mobile networks into following categories:

- Efficient utilization of network resources in CHN environment.
- Technological independent network access, end to end connection and seamless handover.
- Maintaining the certain level of QoS (Quality of Service) for user applications.
- Cooperative network management.
- An intelligent billing policy.

### A. Efficient utilization of network resources in HN environment

The RF spectrum is the scarcest resource in the wireless networks. But this most important and scarcest resource is underutilized in current network standards. If we extend the CHN to the cognitive radios, then one possible way to efficiently utilize the RF spectrum is the opportunistic spectrum access using cognitive radio. In cognitive radio networks the user are divided into two categories primary and secondary users. Primary users are licensed user who can use the licensed portion of RF spectrum while the secondary users can use unlicensed RF spectrum and also attempt to use the unused licensed part of the RF spectrum such that it does not affect the performance of primary users [9]. A MAC level scheduling is also required for optimum spectral utilization in cognitive CHN. This scheduling is different from the conventional MAC level scheduling which aims to give equal distribution of channels among all users. But in cognitive CHN the different users get different channels. The availability of free channels (spectrum holes) varies region to region [13]. In [10] a dynamic spectrum access technique has been proposed for secondary

This work was supported by the National Research Foundation of Korea (NRF) Grant funded by the Korea government (MEST) (no. 2011-0002405).

user to find the spectral holes with minimum spectral collision and overlap time probability. But these techniques will suffer with primary-secondary user spectral collision if secondary user has an inefficient spectrum sensing device.

Spectrum utilization can also be improved by, bypassing the base station if the communicating MTs are near enough to directly send and receive the data, for example 802.11e is defining the mechanism of direct communication between stations with in same Basic Service Set (BSS) area. An Inter BSS Direct Link Setup is proposed in [11], in which a protocol is defined for the direct communication between stations across the BSS. The connection has been set up by using upper layers which makes this direct setup independent of intermediate APs. This has improved the through put about 24 times as compare to the conventional infrastructure. This by-passing technique can be used in cognitive CHN environment if the MTs are close enough to directly communicate.

Moreover, the efficiency of network can be improved by saving the wastage of network resources due to unnecessary network traffic. This can be done by developing the intelligent protocol stack spanning from physical to application level and can build a HN architecture offering the flexible, self aware, self monitoring and robust services.

### B. Technological independent network access, end to end connection and seamless handover

To create a heterogeneous network, the first step is to define a way such that a MT can access different type of networks. UMA/GAN (Unlicensed Mobile Access/Generic Access network) offers a way to access different mobile networks using unlicensed spectrum technologies. But UMA/GAN has some limitations. These limitations has been discussed in [1] and given below.

- For the seamless vertical handover using UMA/GAN the MT should be in the overlapping region of wireless networks.

- UMA/GAN does not support the vertical handover to a better mobile network while a MT moves from one overlapping region to another overlapping region.

- The availability of neigbour list is very important for seamless handover; UMA/GAN does not provide the solution to populate the neigbour list.

- UMA/GAN is the part of 3GPP release 6. This release provides a tightly coupled Network architecture; on the other hand only, flexible network architecture can meet the demands of future networking technologies.

The mobile terminal should be intelligent enough to make best possible decision for selection of network and inter technological switching (selecting the best suitable network), under given circumstances. Therefore, MT should be Reconfigurable Mobile Station having the capability to sense and select the most appropriate network according to the network status and application requirements, by reconfiguring the PHY-layer and tuning the parameters of all upper layers in protocol stack [8]. But this is not enough because, to uniquely identify the MT in CHN environment, to hides the heterogeneity of network from the user and to make technological independent connection, the ene-to-end connection should be IP based connection. Similarly, it is important for a MT to continue its internet session while moving across the HN, this can be achieved by performing IP based handovers. IEEE 802.21 working group has developed a Media Independent Handover (MIH) frame work [2]. This frame work facilitates the protocol stack entities within the node and the network to exchange mobility management information. The frame work also generates and distributes the layer 1 and layer 2 events in a media independent generic format and nodes can exchange the status information of network. But the frame work is incapable to give an acceptable seamless handover mechanism [3], there is a need of further clarification and simplification of commands for seamless handovers to ensure inter-operability among the different handover mechanism (developed by the different service provider independently) running in the heterogeneous environment.

To overcome these limitations, recently different cognitive network architecture projects have been proposed by the researchers, these proposals are independently developed. $E^2R$, m@ANGEL, Sutton at el, CogNet, Thomas at el and SPIN are the most famous. The comparison between these proposals is summarized in [4]. The fundamental concept of these proposals is cognition and active networking but approach of implementation, target services and goals are different. Some of them require high level of support form network element (NE), some of them using centralized approach vs distributed approach for cognitive processes, and some have the ability to reconfigure entire protocol stack vs mid layers and lower layer reconfiguration. But the most important feature is consistency with TCP/IP protocol stack and only the CogNet project funded by NSF is consistent with TCP/IP.

The CogNet is fully distributed and runs the cognitive modules independently at each layer. These modules are inter connected via cognitive bus. CogNet requires moderate level of Network support. However, the goals of CogNet are not completely defined, but due to the consistency with TCP/IP the CogNet is the best choice among all the proposed architectures. To run with the standard protocol, stack the cognitive network elements (CNE) runs cognitive processes in parallel to normal protocol stacks and communicate with each other via cross layer architecture. The cognitive processes analyze the "observed events" of entire protocol stack and exchanges the aggregate results of heterogeneous data with each other and in some cases [5] CNE can stores this spatiotemporal heterogeneous information in a local repository. Similarly, the intelligence can be implant into the data packets, these cognitive packets (CP) has programming codes. The CP executes its code on CNE, shares its experience with CNE and retrieves required information from the local repository.

This stored information is also very helpful for decision making at node and at network level. At node level the cognitive process can use the data of local repository to make decisions, to tune the parameters of entire protocol stack to meet the requirements of running applications and to improve the overall performance of network. This stored data can be used to reduce the latency delay in handovers and make

This work was supported by the National Research Foundation of Korea (NRF) Grant funded by the Korea government (MEST) (no. 2011-0002405).

seamless handover, while ensuring the inter-operability among different networks. A novel intelligent MT architecture is presented in [8]. The MT utilizes the spatio-temporal information to make a proper inter technological switching decision.

The incoming MT can retrieve the aggregated values of protocol stack parameters and can continue the internet session with these parameters, because these parameters are truly reflecting the network status and there is no need to determine appropriate parameters. However in emergency case when some abnormal events detected, the cognitive processes can take independent decision to tune the parameters without querying the repository.

### C. Maintaining the certain level of QoS for user applications

Categorization of applications according to their QoS demand is the first step to ensure the application QoS, but some kind of trade-off always exist between applications QoS and network performance. In [5] the applications are categorized according to their QoS requirements and an intelligent technique has been proposed (for Multihoming Router) to balance the application demand and network performance. The scheme is a six-step process in which root mobile router determine the forwarding/receiving interface on the bases of application demands and network status, but at the cost of processing overhead and increase in network complexity and hence increase in the latency delay.

In cognitive CHN environment the end to end QoS provision is stronger than any other networking environment [12]. In cognitive network all CNE gets the feedback from the cognitive process and they have intelligent decision-making ability therefore, based on cognitive feedback and information of QoS demand of application the CNE can maintain QoS in much better way. But the end to end QoS still a big challenge for cognitive CHNs.

Similarly, in the cognitive CHN environment the MT will be capable of selecting the appropriate network on the basis of QoS demand of starting/running application [8]. If the MT is enabled to maintain the QoS of running application, it will cause the vertical handover if MT detects the more suitable network for the running application. But quick reconfiguration will cause processing overhead and will also generate extra signaling traffic. Therefore, after selection of appropriate network MT should be banned for reconfiguration for some interval of time. Even MT starts using an application which needs more data rate as compare to the capacity of selected network. In this situation the MT can increase the data by using spectrum holes in available network.

### D. Cooperative network management

In cognitive CHN all the networks indirectly affect each other due to the cognitive users. In future the overall network management in CHN environment will depend upon mutual cooperation. The cooperation between different wireless networks can be achieved by the cognition. The cognitive processes give the awareness about the network condition through experience and observation. In [15] middleware architecture has been proposed for context aware networks.

This Middleware architecture has explained the relationship and interdependencies of different context types and gives a flexible service to add and remove the context information and evaluate this information for applications at run time dynamically.

The spatio-temporal information of these experiences and observations are stored in local repositories of CNE. To utilize this information a Cooperative Network Management (CNM) frame work is required to build. But since now there is no concrete CNM frame work has been proposed. Some of the efforts have been carried out to give a generalized CNM architecture for Radio Resource Management (RRM) in CHN [14].

### E. An intelligent billing policy

The billing policy is an important and challenging part of the cognitive CHNs. The businessmen have invested their money in buying the licensed part of the spectrum and therefore they charge the primary users for using their services. But due to the secondary users, the billing in Cognitive CHN can be divided into two parts.

#### 1) Primary user billing

If the billing amount of a mobile subscriber goes to zero it doesn't mean that it cannot continue the call/internet session. Because in future, the MT will have the ability to use the services of different networks and the MT has the ability to switch over to second best available network. Therefore, to prevent the mobile subscriber from call drop or internet session disconnection, it is important to inform the MT about zero balance condition and give some amount of time to switch over to $2^{nd}$ best available network for which the mobile subscriber have sufficient billing amount. If mobile subscriber doesn't have enough money in the billing account to continue as a primary user of any other licensed network, then the MT will try to continue as a secondary user of available networks. In order to avoid the complete network scan. In first step the MT will scan the spectrum to find only those networks for which subscriber's account has enough money to continue the ongoing call/internet session. This billing information of all subscribed networks will be stored in the local repository and MT should update it after end of each calling/internet session.

#### 2) Secondary user billing

If a secondary user MT accesses the spectrum hole of the licensed spectrum part, then the billing policy depends upon the agreement between the network owners. But to preserve the main advantage of opportunistic DSA the billing charge for secondary should be very low and secondary user should be charged if it affects the performance of primary users. The network will inform the secondary user to release the spectrum hole via common channel or via some centralized controlling entity.

### III. FUTURE NETWORK ARCHITECTURE

This section presents a possible Network Architecture for cognitive CHN. Which helps in improving the efficiency of the networks, tries to maintain the QoS without affecting the Primer user, helps in network management and also give solutions to the billing problems. Figure-1 presents typical cognitive CHN environment.

This work was supported by the National Research Foundation of Korea (NRF) Grant funded by the Korea government (MEST) (no. 2011-0002405).

- The MT-1 is the primary user of Network-4 but it is also using the spectral hole of Network-1,2 and 3 to increase the data rate and to maintain the QoS.

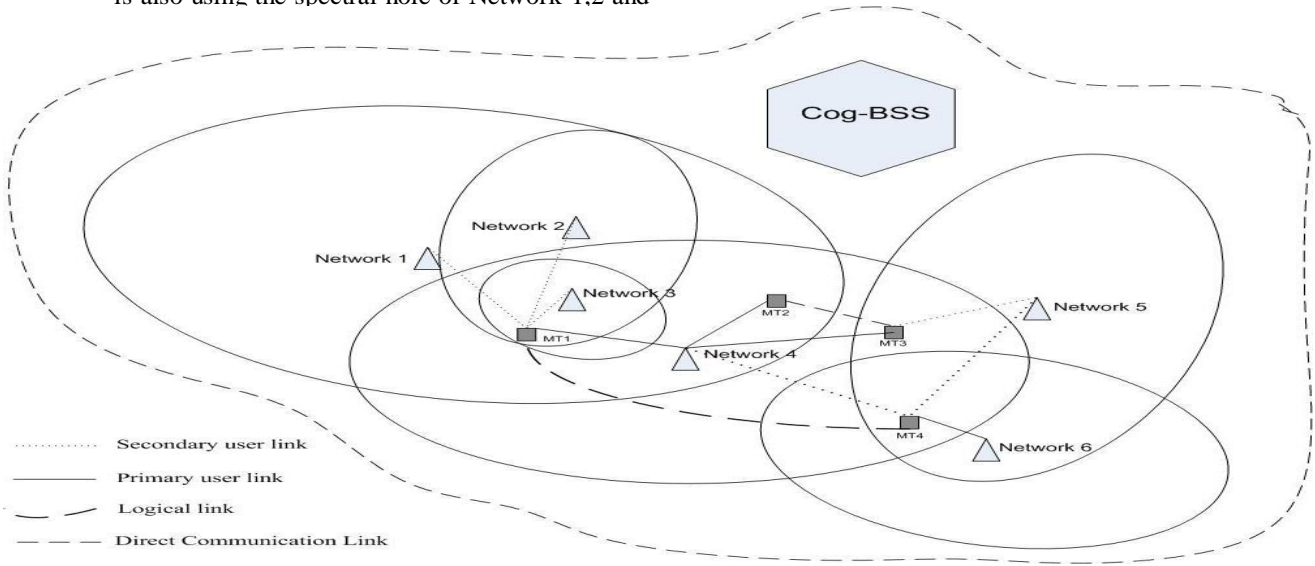

**Figure 1- Cognitive CHN environment**

- MT-2 and 3 are the primary user of Network-4 but both are close enough to send and receive data directly therefore they communicate directly after initial connection setup via Base station (IEEE802.11e is using the same technique to improve the spectrum usage efficiency). The MT-3 is also using the spectral hole in Network-5.

- The MT-4 is the primary user of Network-6 but it is also using the spectral hole of Network-4 and 5 to increase the data rate and to maintain the QoS. MT-1 and MT-4 both are communicating with each other and the curve shows the logical connection between MT-1 and MT-4.

- The Cognitive Basic Service Set (Cog-BSS) is an independent entity. It performs the critical role to make sure that the secondary users do not affect the performance of primary users.

A. *Mobile Terminal*

In Cognitive CHN the MT will play a crucial role. To make the network side simpler and to prevent the network side form large scale, it is important to make MT more flexible and powerful in processing. Based on the work by B.S. Manoj et al [4b], in [8] the MT architecture for cognitive CHN has been presented. The MT uses the observed knowledge of protocol stack stores in local repository and makes the decision to optimally use the network resources. The main focus is to make proper and in time inter technological switching by estimating and predicting the channel conditions.

The figure-2 shows the MT architecture. This architecture is using the distributed cognitive module approach presented by B.S. Manoj et al [4b]. Each layer has its own cognitive module; all are inter connected via cross layer bus. From the architectural point of view it is better to divide the repository into five different partitions, reserved for each layer. The cognitive modules store the spatiotemporal information in the allocated memory space of repository. The Logic Analyzer can request the required information from local repository and if something critical situation occur Logic Analyzer can also get directly input from layers. The decision Maker tunes the protocol stack on the basis of input from logic analyzer and spectrum scanner. But for fast and seamless inter technological switching the Decision Maker can directly send the instruction to Reconfiguration memory to transfer the PHY layer logic and MAC layer parameters.

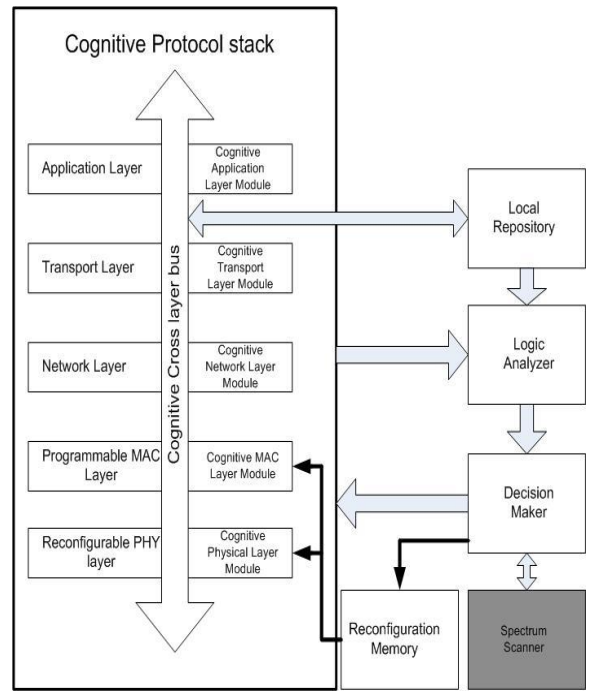

**Figure 2- MT Architecture**

This work was supported by the National Research Foundation of Korea (NRF) Grant funded by the Korea government (MEST) (no. 2011-0002405).

The operation of the MT in cognitive CHN is shown in figure-3. The flow chart shows that the MT do its best effort to increase the data rate and improve the QoS of applications but at the same time it releases the spectral hole if it receives channel release information for a particular network. This immediate release of channel is necessary to prevent the primary user from suffering bad network situation.

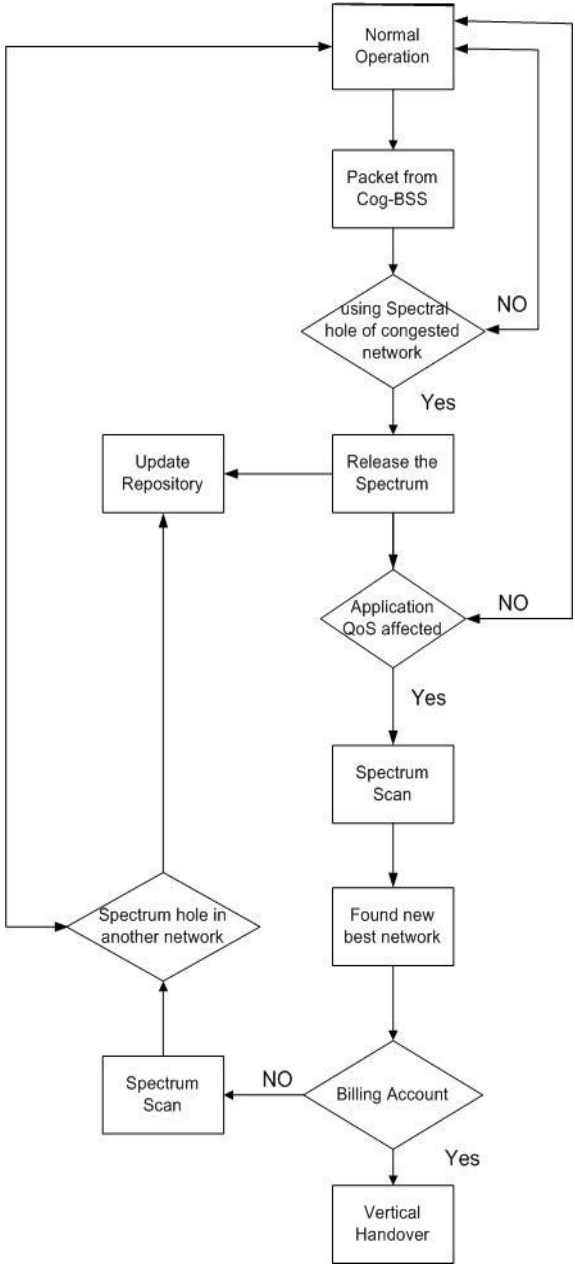

**Figure 3- Operation flow chart**

B. *Cognitive Basic Service Set (Cog-BSS)*

The Cog-BSS is directly connected to all networks via common interface and it sends the instruction to all secondary users on a common broadcast channel. It consists of four major components, as shown in figure-4,

1) *common interface to Networks.*
2) *Information Analyzer.*
3) *Instructor.*
4) *Common broadcast Channel.*

Cog-BSS can receive the control packets from all networks in its coverage area. This control packet consists of congestion information of network and aggregate performance of primary users with information of minimum acceptable performance of primary users. The "Information Analyzer" makes analysis of the information received from the network and tells the "Instructor" to issue appropriate command to the primary users. For example, received information shows that the congestion situation of Network-2 is going worst. Therefore "Instructor" will issue "Release-channel" command to the all secondary users of Network-2 on common broad cast channel.

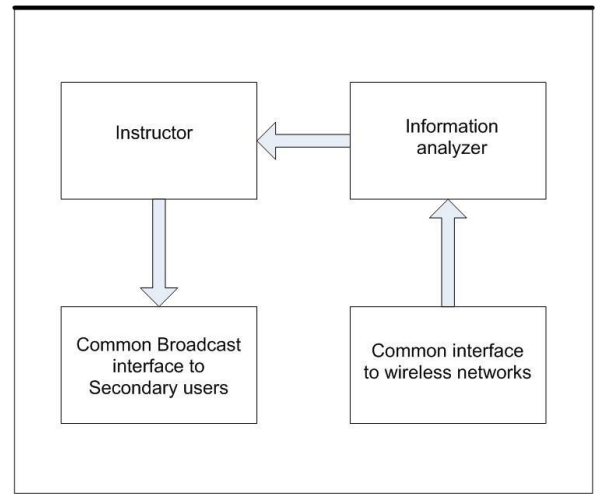

**Figure 4- Cognitive Basic Service Set**

IV. CONCLUSION

The current state of art is not fulfilling the future network requirements and recent research on cognitive radios has extended to complete protocol stack. The introduction of cognition in upper layer has made the things simpler and easier and has provided a foundation for next generation networks. But still a lot of research is required to give a complete unique CHN architecture. There is a need to introduce some algorithms schemes and mechanism to resolve the problems of quick and secure host identification and a quick and seamless handover mechanism in the mobile environment, the application demands for QoS is varying application to application, an intelligent scheme is required to balance the application demand and network performance without wastage of network resources and without compromising a certain level of QoS, the Cooperative Network Management (CNM) frame work is required to build and to attract the business community an efficient billing strategy for secondary user should be introduce. The network architecture proposed in this paper can handle all these issues but the architecture is giving an abstract concept and further in-depth explanation is required. Especially the use of new entity "Cog-BSS" can further be explained to use it in implementation of Cooperative Network Management (CNM) frame work. The stored information in local repositories of MT will play a crucial role in future mobile

This work was supported by the National Research Foundation of Korea (NRF) Grant funded by the Korea government (MEST) (no. 2011-0002405).

network. There is a need to provide a sophisticated Database system that can fulfill requirements to perform quick spatiotemporal queries.

This work was supported by the National Research Foundation of Korea (NRF) Grant funded by the Korea government (MEST) (no. 2011-0002405).